Using Direct Policy Search to Identify Robust Strategies in Adapting to Uncertain Sea-Level Rise and Storm Surge

Gregory G. Garner[1]

Klaus Keller[2,3,4]

[1] Woodrow Wilson School of Public and International Affairs, Princeton University, Princeton, NJ, 08540, USA

[2] Department of Geosciences, The Pennsylvania State University, University Park, PA, 16802, USA

[3] Earth and Environmental Systems Institute, The Pennsylvania State University, University Park, PA, 16802, USA

[4] Department of Engineering and Public Policy, Carnegie Mellon University, Pittsburgh, PA, 15213, USA

**ABSTRACT**

Sea-level rise poses considerable risks to coastal communities, ecosystems, and infrastructure. Decision makers are faced with uncertain sea-level projections when designing a strategy for coastal adaptation. The traditional methods are often silent on tradeoffs as well as the effects of tail-area events and of potential future learning. Here we reformulate a simple sea-level rise adaptation model to address these concerns. We show that Direct Policy Search yields improved solution quality, with respect to Pareto-dominance in the objectives, over the traditional approach under uncertain sea-level rise projections and storm surge. Additionally, the new formulation produces high quality solutions with less computational demands than an intertemporal optimization approach. Our results illustrate the utility of multi-objective adaptive formulations for the example of coastal adaptation and point to wider-ranging application in climate change adaptation decision problems.

**SOFTWARE AVAILABILITY**

Model source code and data are available at https://doi.org/10.18113/D3XD32. Model requires Gnu C++ compiler 5.3.1 (https://gcc.gnu.org/), OpenMPI 1.10.1 (https://www.open-mpi.org/), NetCDF 4.4.1 (https://www.unidata.ucar.edu/software/netcdf/), Boost 1.61.0 (http://www.boost.org/), and Borg 1.8 (http://borgmoea.org/) or later versions.

**1. INTRODUCTION**

Sea-level rise (SLR) drives considerable risks to coastal communities, ecosystems, and infrastructure around the world (Eijgenraam et al., 2014; Le Cozannet et al., 2015; Miller et al., 2015; Moftakhari et al., 2015; Nicholls and Cazenave, 2010). The Intergovernmental Panel on Climate Change reports that global mean sea-levels will likely rise by 0.52 to 0.98 m by the year 2100 (relative to the 1986-2005 period) under a high greenhouse gas concentration scenario (Church et al., 2013) and that a collapse of portions of the Antarctic ice sheet would irreversibly drive SLR well beyond this range (DeConto and Pollard, 2016; Pollard et al., 2015; Wong et al., 2017; Wong and Keller, 2017). Additionally, climate change is contributing to changes in the distribution of storm surge events, especially with regard to the extreme tail-area events (Arns et al., 2017; Grinsted et al., 2013, 2012; Neumann et al., 2015). Though SLR and storm surge have been, and continue to be, extensively studied, they remain deeply uncertain across decision-relevant time scales (Buchanan et al., 2016; Hinkel et al., 2014; Le Cozannet et al., 2015; Lempert et al., 2004; Lempert et al., 2012). Those in position to enable coastal adaptation strategies rely on decision support tools to process this deeply uncertain information to inform their decisions (Lempert et al., 2004; Lempert et al., 2012; Liverman et al., 2010).

Developing and applying these decision support tools poses conceptual and methodological challenges. One approach is to build an optimization tool that finds the time-series of dike heighteneings that minimizes the total economic cost of building dikes or levees (Eijgenraam et al., 2014; Kind, 2014; Slijkhuis et al., 1997; Speijker et al., 2000; Van Dantzig, 1956; van der Pol et al., 2014). To incorporate uncertainty, this process can be repeated over various sets of model parameters and the expectation of the total costs can be minimized.

This approach, however, is silent on several key aspects of decision making. First, the single-objective formulation can hide important tradeoffs among stakeholder preferences of which the decision maker must be aware (Garner et al., 2016; Quinn et al., 2017; Singh et al., 2015). For example, a climate mitigation strategy derived by maximizing the expectation of an *a priori* defined utility function may be blind to important tradeoffs in environmental objectives and remove relevant stakeholders from the negotiations (Garner et al., 2016). Second, insufficient sampling of uncertainty can under-represent extreme events that may weigh heavily in the decision (Garner et al., 2016; Lempert et al., 2004). Lastly, this formulation does not make use of important state-related information, such as the level of the water with respect to the top of the dike, that can be used to inform the decision (Quinn et al., 2017; van der Pol et al., 2014). The Robust Decision Making (RDM) framework provides a means of approaching these concerns (Herman et al., 2015; Kwakkel et al., 2016; Lempert et al., 2006; Weaver et al., 2013). We expand on this framework with an additional component to include endogenous learning and adaptive decision making.

In this study, we reformulate the problem to begin addressing these concerns. Specifically, we split up the total cost metric into its investment cost and damage components to illustrate the direct tradeoffs between the two objectives. We use a states-of-the-world (SOWs) approach about SLR and storm surge to introduce uncertainty to the SLR adaptation model and provide coverage of tail-area events. Finally, we apply Direct Policy Search (DPS), an adaptive state-based method of endogenous learning, to incorporate new information and adapt the decision through the simulation period (Deisenroth et al., 2013; Giuliani et al., 2016). We hypothesize that these changes will provide an improvement in solution quality over the traditional approach.

## 2. METHODS

The following sub-sections describe our approach to formulating the problem and designing the experiment. The sub-sections largely follow the taxonomy proposed in the XLRM framework where the decision problem is comprised of exogenous uncertainties (X), levers or actions at the disposal of the decision maker (L), the model or relationship (R) mapping the decision maker's actions to the performance metrics or objectives (M) (Lempert et al., 2006).

### 2.1 Base Model (R)

The base model used in this analysis is an SLR adaptation model used in the Netherlands to help inform safety standards for the numerous dikes protecting the country. The model is described extensively in (Eijgenraam et al., 2014). The key components are briefly summarized below.

The objective is to find the time series of annual dike heightenings $u_t$ that minimizes the total discounted social cost over the simulated time horizon of 300 years

$$\min \left\{ \sum_t I(h_t^-, u_t) e^{-\delta t} + S_t e^{-\delta_1 t} \right\}, \quad (1)$$

where $I$ is the investment cost to heighten the dike and $S_t$ is the expectation of damages at year $t$. Both investment cost and the expectation of damages are discounted by a factor of $\delta$ and $\delta_1$ respectively. The investment cost component is defined by an exponential function of the increase in dike height at a given time

$$I(h_t^-, u_t) = \begin{cases} 0 & if\ u_t = 0 \\ (\kappa + v u_t) e^{\lambda(h_t^- + u_t)} & if\ u_t > 0 \end{cases}, \quad (2)$$

where $\kappa, v,$ and $\lambda$ are positive constants and $u_t$ is the additional height added at time $t$ to the dike at height $h_t^-$. The increase in dike height reduces the probability of a flood ($P_t$) and thus reduces the expectation of damages according to

$$S_t = P_t V_t \quad (3)$$

$$V_t = V_0^- e^{\gamma t} e^{\zeta(H_t - H_0^-)}, \quad (4)$$

where $V_t$ is the damage incurred in the event of a flood at time $t$, $V_0^-$ is the damage incurred by a flood before $t = 0$, $H_t$ is the dike height at time $t$, $H_0^-$ is the dike height just before $t = 0$, $\gamma$ is the economic growth rate within the area protected by the dike, and $\zeta$ is the increase in loss per unit of dike heightening. Values for these parameters are listed in Table 1, which are consistent with the parameter values for dike ring 16 in (Eijgenraam et al., 2014). The probability of flooding ($P_t$) is handled differently in our formulations and is discussed in section 2.2.

| Parameter (symbol) | | Value | Unit |
|---|---|---|---|
| Discount rate of investment cost | ($\delta$) | 0.04 | yr$^{-1}$ |
| Discount rate of expected damages | ($\delta_1$) | 0.04 | yr$^{-1}$ |
| Initial investment cost to heighten dike | ($\kappa$) | 324.6287 | Euros |
| Linear parameter in investment cost | ($\nu$) | 2.1304 | Euros/yr |
| Exponential parameter in investment cost | ($\lambda$) | 0.01 | cm$^{-1}$ |
| Economic growth rate within dike | ($\gamma$) | 0.02 | yr$^{-1}$ |
| Increase in loss per unit of dike heightening | ($\zeta$) | 0.002032 | cm$^{-1}$ |
| Initial height of dike prior to t=0 | ($H_0^-$) | 118.6837 | cm |
| Loss due to flood prior to t=0 | ($V_0^-$) | 22656.5 | Euros |

Table 1. Parameter values of the economic component of the base model

## 2.2 Uncertainty in SLR and Storm Surge (X)

In the base model, the probability of flooding is represented by an exponential distribution of extreme flood events. A steady rate of increase in the effective water height is used to represent rising sea-levels, and in the case of the Netherlands, land subsidence. This rate parameter is used in the exponential distribution to determine the probability of a dike failure as a function of time. Sea-level rise, however, is a deeply uncertain consequence of a changing climate (Church et al., 2013) and a steady rate of sea-level rise represents only one possible future state. In the reformulated analysis, the probability of flooding is replaced with an explicit representation of states of the world (SOWs) (Garner et al., 2016; R. J. Lempert et al., 2012; Singh et al., 2015). In this approach, parameters that represent future states of the world do not have a single value, but rather a distribution of possible values. Drawing a sample from each parameters' distribution would represent a single state over which the model is evaluated. Repeating this process provides a series of outcomes from which expectations and reliability metrics can be calculated.

In order to use the SOW approach to represent uncertainty, we incorporate new structural representations of sea-level rise and storm surge events into the base model. Future mean annual sea-level rise is approximated by the approach used in Lempert et al. (2012)

$$z_t = \begin{cases} a + bt + ct^2 & if\ t \leq t^* \\ a + bt + ct^2 + c^*(t - t^*) & if\ t > t^* \end{cases}, \qquad (5)$$

where parameters $a$, $b$, and $c$ are the initial sea-level rise anomaly, linear rate of change of sea level, and the acceleration of sea-level change respectively. The $c^*$ and $t^*$ parameters represent a potential abrupt change in sea-level rise such as the sudden collapse of an ice sheet (DeConto and Pollard, 2016; Pollard et al., 2015). The linear rate would increase by $c^*$ when $t$ exceeds $t^*$ in the simulation. The joint distribution of these parameters are estimated through the calibration process described in Oddo et al. (2017) and used in this analysis to derive SOWs.

Storm surge events occur on top of the mean annual sea level. These events are estimated through inverse-transform sampling of the stationary generalized extreme value (GEV) distribution calibrated in Oddo et al. (2017).

$$x_t = \begin{cases} \mu + \sigma \ln\left(\frac{1}{\ln(1/p)}\right) & if\ \xi = 0 \\ \mu + \frac{\sigma\left((\ln(1/p))^{-\xi}\right)}{\xi} & if\ \xi \neq 0 \end{cases}, \qquad (6)$$

where $\mu$, $\sigma$, and $\xi$ are the location, scale, and shape parameters of the GEV function and $p$ is a quantile randomly sampled from a uniform distribution from zero to one inclusive. A draw from this distribution represents an annual maximum surge event ($x_t$) that is added to the mean annual sea level ($z_t$) at time step $t$. Note that this distribution is stationary and does not account for changes in the maximum surge event over time. Additionally, the distribution was calibrated to observations with the annual mean sea-level signal removed. Tides, wave set up, and wave run up contribute to extreme water levels, but were not addressed in the calibration process. Estimates of the joint distribution of the parameters in equation 6 from Oddo et al. (2017) are used in this analysis to derive SOWs.

To represent a large portion of the tails of the distributions of this uncertainty, this analysis produces 100,000 SOWs. We evaluate the model over each SOW to produce the objective values described in section 2.3.

## 2.3 Objectives (M)

The original objective in the base model is to minimize the total social cost (eq. 1). The total social cost is the discounted sum of two components, the total investment cost of building up the dike height (eq. 2) and the total expectation of damages (eq. 3). While this seems like a reasonable objective, it neglects the distinct possibility that multiple decision makers, representing diverse stakeholder preferences, may be involved in the negotiations. For example, some stakeholders are so risk-adverse that the decision makers, acting on their behalf at the negotiations, are willing to invest heavily into infrastructure in order to minimize expected damages (Eijgenraam et al., 2014; Kind, 2014) while decision makers representing fiscally-conservative stakeholders would argue for low investment-cost solutions. In either case, a traditional single-objective function masks these tradeoffs leading to a single optimal solution that could be inconsistent with these stakeholders' preferences (Garner et al., 2016).

Instead of combining the two components of the total social cost into a single metric, this analysis defines the two components as separate objective functions, both of which are to be minimized

$$Objective\ 1: min\left\{\sum_t I(h_t^-, u_t)e^{-\delta t}\right\} \tag{7}$$

$$Objective\ 2: min\left\{\sum_t S_t e^{-\delta t}\right\}. \tag{8}$$

Note that these objectives use the same discount factor $\delta$ as opposed to the separate discount factors used in the objective function in the base model (eq. 1). Additionally, the probability of the water level exceeding the dike height at time $t$ used in calculating the expected damages ($P_t$ in eq. 3) is replaced with the proportion of SOWs with water levels that exceed the dike height at that time. The result is, instead of a single optimal build policy, a set of Pareto-optimal solutions for which it is not possible to find another solution that reduces investment (damage) costs without increasing damage (investment) costs.

### 2.4 Decision Variables – Intertemporal and Direct Policy Search (L)

One straight-forward approach to this optimization problem is to treat the heightening at every time-step in the simulation as a decision variable. With a time-horizon of 300 years and an annual time-step, this results in an intertemporal optimization with 300 decision variables. In this approach, the heightening at each time-step is exogenously prescribed to the model.

One important drawback of this approach is that the same build policy is used regardless of the state of the system. For example, a specific policy that is appropriate for a rapid increase in water height would unnecessarily heighten the dike if the water levels rise slowly or not at all. Conversely, a policy that is appropriately built for slow-rising water levels would be catastrophically underprepared if the water levels rise abruptly and quickly. Additionally, this approach can be a poor approximation of how adaptation decisions are made in real-world situations. If a dike fails or is clearly in danger of failing, action would be taken to replace or strengthen the dike outside of the planned heightening policy in place. In other words, new observations provide information about the system that can be used in further decisions. This learning process is absent from the intertemporal approach, but can be included using Direct Policy Search (DPS) (Deisenroth et al., 2013; Giuliani et al., 2016; Quinn et al., 2017).

In the DPS approach, the decision whether or not to heighten the dike in any given time-step depends on the state of the system. A state variable is defined as some observable metric that relates back to the current state of the system. A function then maps the state variable or variables to one or more signposts that trigger actions. For example, the proximity of the water level with respect to the top of the dike may trigger a heightening of the dike to prevent overtopping. Since DPS is state-dependent, applying it to a model with uncertainty should reduce unnecessary heightenings in slow SLR states and be robust to situations of abrupt acceleration of SLR.

Our analysis considers two state variables for the DPS formulation, the observed mean annual rate of change in the water level at the dike ($\beta_t$) and the square root of the sum of squared residuals ($srss_t$) between the observed water level and fitted water levels derived from a linear model regressed on the previous 30 simulation years of water level (mean annual SLR and surge) at time t. Let $i = \{1,2,\ldots,30\}$ represent the 30 simulation-year window prior to time t over which a linear model is regressed (i.e. when $t = 2100$, $i = 1$ refers to $t = 2070$ and $i = 30$ refers to $t = 2099$). The linear model is

$$\hat{y}_i = \alpha_t + \beta_t i, \tag{9}$$

where $\hat{y}_i$ is a fitted water level height at the $i^{th}$ position in the 30 simulation-year window, $\alpha_t$ is the fitted intercept and $\beta_t$ is the fitted slope.

$$\beta_t = \frac{\sum i y_i - \frac{1}{n}\sum i \sum y_i}{\sum i^2 - \frac{1}{n}(\sum i)^2} \quad ; \quad \alpha_t = \bar{y} + \beta_t \bar{i}. \tag{10}$$

The value $\bar{y}$ is the mean of the water levels in the $n = 30$ simulation-year window and $\bar{i}$ is the mean of set $i$. The value $y_i$ is the water level at the $i^{th}$ position in the 30 simulation-year window ($y_i = z_{t-n+i} + x_{t-n+i}$; eqs. 5,6). Each of the summations integrate from $i = 1$ to $n$. The $srss_t$ is thus defined as

$$srss_t = \sqrt{\sum_i (\hat{y} - y_i)^2}. \tag{11}$$

Each of these state variables contributes to the two quantities that determine the action for the associated time-step. The first quantity is the buffer height, which acts as the signpost and is defined as the minimum distance between the water level and the top of the dike that is considered safe. If the difference between the water level and the height of the dike is less than the buffer height, action is triggered and the dike is heightened. The second quantity is the freeboard height, which is the additional heightening built on top of the minimum heightening that would bring the dike back to a safe height. A freeboard height is often applied (Davis et al., 2008; Gui et al., 1998) to reduce the probability of needing dike heightenings in consecutive years since there is an initial cost to starting the heightening process (parameter c in eq. 2) and heightening in consecutive simulation years would be expensive.

The state variables map to the buffer height and freeboard height through quadratic relationships

$$BH_t = v_1\beta_t^2 + v_2\beta_t + v_3 srss_t^2 + v_4 srss_t + v_5 \tag{12}$$

$$FH_t = v_6\beta_t^2 + v_7\beta_t + v_8 srss_t^2 + v_9 srss_t + v_{10}, \tag{13}$$

where $v_j$ are the decision variables provided by the optimization process. At time t, the freeboard and buffer heights are calculated and a heightening for time t is determined.

$$u_t = \begin{cases} 0 & if\ BH_t < h_t^- - y_t \\ y_t - (h_t^- - BH_t) + FH_t & if\ BH_t \geq h_t^- - y_t \end{cases}, \tag{14}$$

For example, if the water level is 30 cm, the dike height is 40 cm, the buffer height is 15 cm, and the freeboard height is 10 cm, then a heightening of the dike would be triggered (15 cm > 40 cm – 30 cm), the dike would be heightened by the amount needed to achieve the buffer height (5 cm) plus the freeboard height (10 cm) resulting in a final dike height of 55 cm.

For comparison, we discuss the results from both an intertemporal (non-adaptive) approach and the DPS (adaptive) approach described above in the following sections. The objectives, uncertainty, and other modeling aspects are consistent between the two approaches.

**2.5 Optimization**

We use the Master-Slave Borg Multiobjective Evolutionary Algorithm (MS Borg-MOEA) (Hadka and Reed, 2012) to optimize the system and identify the Pareto-optimal solution set for both the intertemporal and DPS formulations. Evolutionary algorithms are common in optimization. The basic idea is to propose a set of possible solutions, select a set of well performing solutions based on the values of the objective functions, and mutate these solutions to produce the next set of proposed solutions. This process continues until a set time or number of function evaluations has been performed. We briefly describe some of the key features specific to the MS Borg-MOEA that are pertinent to this problem below.

First, MS Borg-MOEA auto-adapts its search based on the level of performance each of its six operators achieves with respect to producing improvements in the solution archive. Search operators that produce large Pareto improvements in the solution archive are given more weight in subsequent iterations. This allows the algorithm to quickly navigate to and converge upon the Pareto-front in objective space. Second, the epsilon-dominated archive allows user-defined resolution in objective space and guarantees solution diversity by requiring that a solution provide a Pareto improvement no less than a threshold (epsilon) change in the objective values. These were desired features given the complexity of the two problem formulations in this study.

Evolutionary algorithms, including MS Borg-MOEA, rely on random number generation. The choice of seed value to the random number generator is critical to the performance of the optimization algorithm. It is possible that, in a predefined number of function evaluations, two seed values can lead to very different results. To mitigate this problem, the optimization process is repeated 50 times, each with a unique seed value passed to the random number generator in MS Borg-MOEA. Results from each of the 50 runs are combined and sorted to find the Pareto-optimal solution set.

The experiment is summarized in Table 2. Aside from the number of decision variables in each of the formulations (300 for the intertemporal formulation and 10 in the DPS formulation), both formulations consisted of the same base model, SOWs, objectives, and number of random number generator seeds to test in the optimization process. Runtime dynamics, consisting of operator selection probabilities, archive snapshots, wall time, and number of function evaluations, were captured at every 200 function evaluations. We require solutions to be somewhat consistent with practice in that no adaptation solution should perform poorly with respect to reliability. However, imposing too high of a reliability constraint would make it difficult to find feasible solutions within the allotted function evaluations of the experiment. To achieve a balance between these points, we apply a relatively loose reliability constraint where the adaptation solution must achieve at least 80% reliability over time and across SOWs.

|  | Direct Policy Search | Intertemporal |
|---|---|---|
| **Decision Variables** | 10 ($v_1, v_2, ..., v_{10}$; Eqs. 12,13) | 300 ($u_1, u_2, ..., u_{300}$) |
| **Objectives** | Investment cost (min), Expected damages (min) ||
| **Constraints** | 80% reliability over time and SOWs ||
| **Uncertainty** | Sea-level rise and storm surge (100k SOWs) ||
| **Optimization Iterations** | 50 seed values, 200k function evaluations, snapshots every 200 function evaluations ||

Table 2. Experiment implementation. Both the Direct Policy Search and Intertemporal formulations use the same objectives, constraints, uncertainty, and optimization settings. The only difference is the type and number of decision variables (row 1).

## 3. RESULTS

The DPS formulation produces solutions that completely dominate the intertemporal formulation with respect to investment cost and expected damages (Fig. 1). Our implementation of the intertemporal formulation (magenta) is unable to produce a solution with investment costs below €251 million whereas 90% of our DPS formulation solutions produce investment costs below €250 million. Furthermore, the expected damages incurred with the minimum investment cost solution from the intertemporal formulation is over three orders of magnitude greater than the expected damages from the DPS solution of equal investment cost (intertemporal - €29.7 billion; DPS - €11.9 million). The minimum cost solution, akin to the objective of the base model, is four times greater in the intertemporal formulation compared to the DPS formulation (intertemporal – €972 million; DPS – €236 million), largely due to the difference in investment cost.

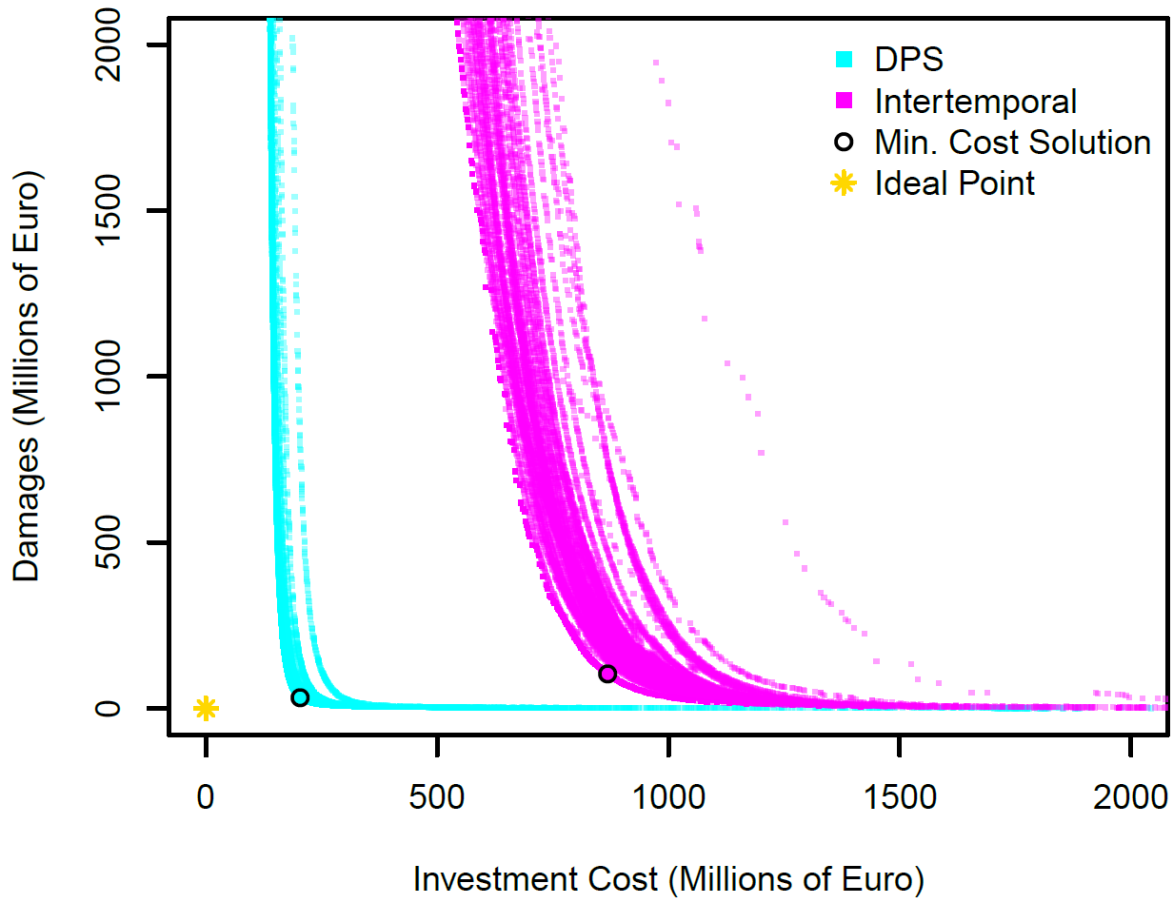

Figure 1. Tradeoff curves (Pareto-fronts) of the intertemporal formulation (magenta) and the DPS formulation (cyan). Points along each of the curves are solutions that cannot improve in one objective without deteriorating in another. The ideal solution (gold star) is a solution that produces zero investment costs and zero expected damages. The minimum total cost solution (black circle) is the solution that minimizes the sum of the discounted investment costs and expected damages akin to the objective of the base model.

The added complexity in the DPS formulation adds additional computation time to a single function evaluation when compared to the intertemporal formulation; however, solutions of the same relative quality are found in fewer function evaluations in the DPS formulation than in the intertemporal formulation (Fig. 2). Hypervolume is a measure of Pareto-dominated volume in objective space relative to a reference solution set. Fig. 2 illustrates the evolution of the Pareto-front through the optimization process for both the DPS (cyan) and intertemporal (magenta) formulations. A hypervolume of 0% indicates that no solutions have been found while a hypervolume of 100% indicates that the Pareto-front being tested dominates the same volume of objective space as the reference solution set. In this analysis, the reference set to which each formulation is assessed is the overall Pareto-front across both formulations. The DPS formulation begins finding solutions an order of magnitude faster than the intertemporal. Every seed tested in the DPS formulation was able to find at least one solution within 200 function evaluations and a tradeoff between the objectives (2 or more solutions) within 2000 function evaluations. In contrast, the intertemporal formulation required 3600 function evaluations to find a solution and 6800 function evaluations to find a tradeoff in any of the tested seeds. When the intertemporal formulation begins finding solutions, the DPS formulation has already covered over 98% of the volume of the reference solution set. Finally, when the DPS formulation comes within 1% of the volume of the reference solution set, the intertemporal formulation has covered only approximately 11.5% of the reference solution set. While we do not quantify convergence of solutions within a specific formulation, the spread in the intertemporal hypervolume obtained by the final function evaluations across the seeds suggests that many of the seeds tested in the intertemporal formulation may not have converged over the 200,000 function evaluations allotted in the experiment.

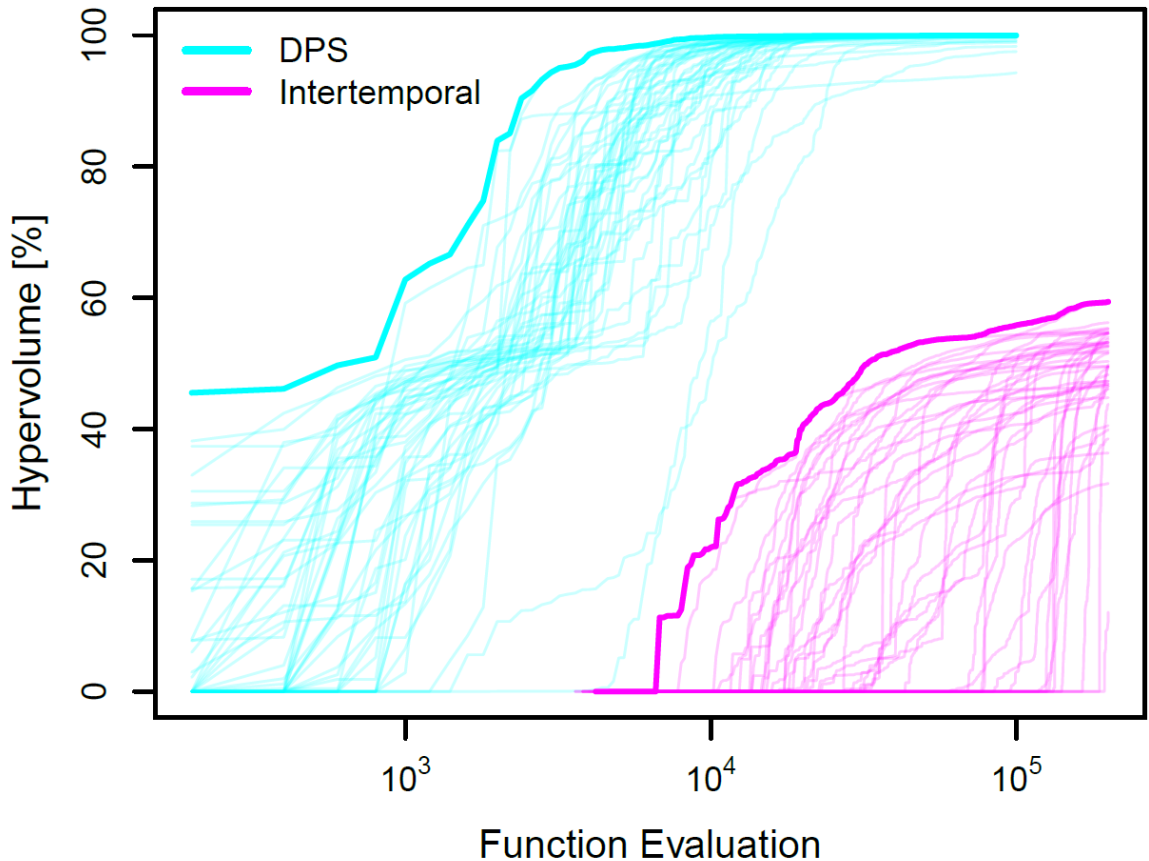

Figure 2. Evolution of the Pareto-front hypervolume during the optimization process for the DPS (cyan) and intertemporal (magenta) formulations. Hypervolume is a metric of the volume of objective space dominated by a specific Pareto-front. In this application, zero hypervolume indicates no feasible solution has been found and 100% hypervolume means the solution archive has converged onto the overall Pareto-front combined from both experiments shown in Fig. 1. Thick-dark lines are the best solutions for each formulation when combining all 50 runs. Thin-light lines are individual runs.

The drastic difference between the two formulations, with respect to the minimum total social cost solution, can be explained by the build policies generated by the formulations (Fig. 3). The minimal total cost solution from the intertemporal formulation produces a single build policy that is independent of the SOW (black line) whereas the minimal total cost solution from the DPS formulation adjusts the build policy depending on the observed state. Note that for any given time, the observed state is derived from the prior 30 years of the water level in a specific SOW (see section 2.4). For example, decision to build in year 2150 depends on the SOW, the height of the dike, and the water level from the years 2120 – 2149. As a result, the intertemporal formulation, in order to satisfy the strict reliability constraint, is required to considerably heighten the dike over time to prevent overtopping in the most extreme water level cases. The DPS solution reduces the investment costs by reacting to the observed state and building appropriately. In the high-water level example (red), the build policy is slightly more aggressive than the intertemporal solution, but for the low and medium water level examples (green and blue, respectively) the build policy relaxes, building less frequently and in shorter height increments over the simulation. The magnitude and the rate of the heightenings are also much more consistent in the DPS formulation compared to the intertemporal.

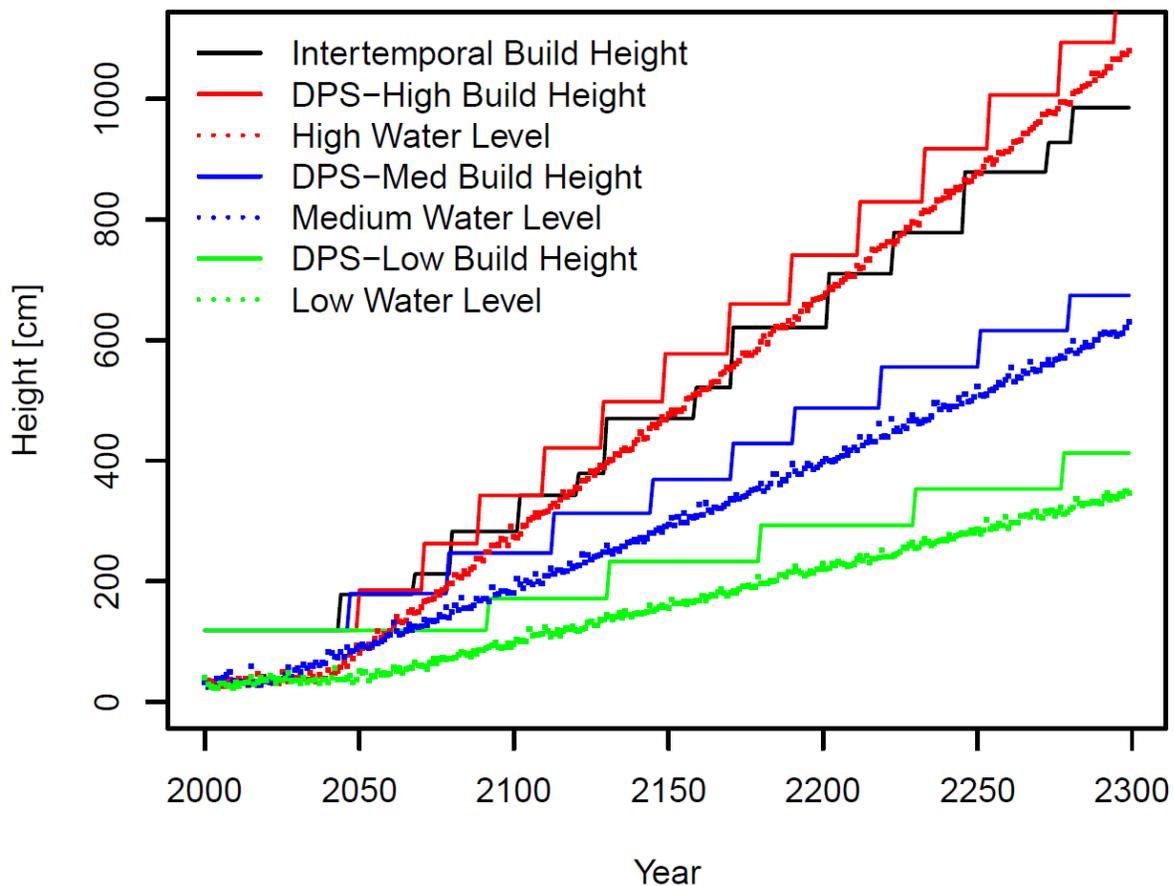

Figure 3. Time series of dike height for the minimum total cost solution for the intertemporal formulation (solid black line) and DPS formulation (solid color lines). The water level (color dots) for three different SOWs (low – green, med – blue, high – red) are provided as example observations that are used to derive the buffer and freeboard heights for their respective

SOWs. The color of the dike height for the DPS formulation matches the color of the water level for that specific SOW.

The state-varying freeboard and buffer heights enable the DPS formulation to Pareto-dominate the intertemporal formulation (Fig. 4, 5). The largest advantage to this method is in identifying the threshold event. The state variables used in the DPS formulation are able to provide a signal that produces a shift in both the freeboard height and the buffer height (Fig. 4). The freeboard height, largely controlled by the $srrs_t$ (Fig. 5), tends to decrease approximately 10 cm across the time of the threshold event and the variance approximately triples. The buffer height, also largely influenced by the $srrs_t$ (Fig. 5), follows an inverse pattern where the height increases by approximately 8 cm across the threshold event, but the variance remains the same before and after the threshold event.

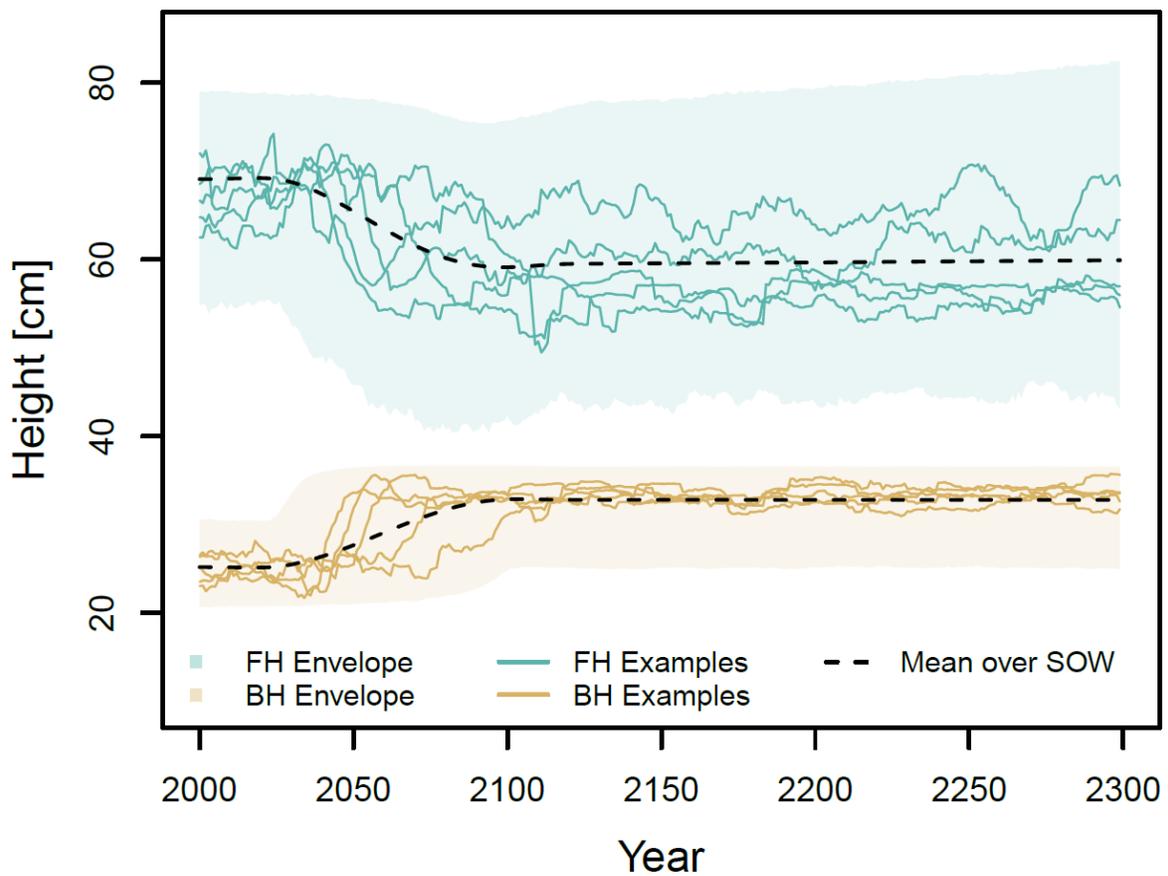

Figure 4. Calculated freeboard height (blue) and buffer height (orange) from the minimum total cost solution from the DPS formulation. The dotted line and shaded region represent the mean and range respectively across the SOWs. The solid color lines represent a time-series of freeboard and buffer heights from randomly-selected SOWs.

While the DPS formulation produces superior solutions when compared to the intertemporal formulation with respect to the objectives, there exist a small percentage of individual SOWs where the DPS formulation falls short of the intertemporal formulation. Of the 100,000 SOWs in this analysis, the intertemporal formulation outperforms the DPS formulation in 0.668%, 1.366%, and 1.054% of the SOWs with respect to investment costs, damages, and total cost respectively in their respective minimum total cost solutions. Additionally, each of the SOWs where the intertemporal formulation outperforms the DPS formulation in total costs can be found in the set of SOWs for the investment cost (179) and damages (882). There are seven SOWs where both the investment cost and damages are less than those in the DPS solution.

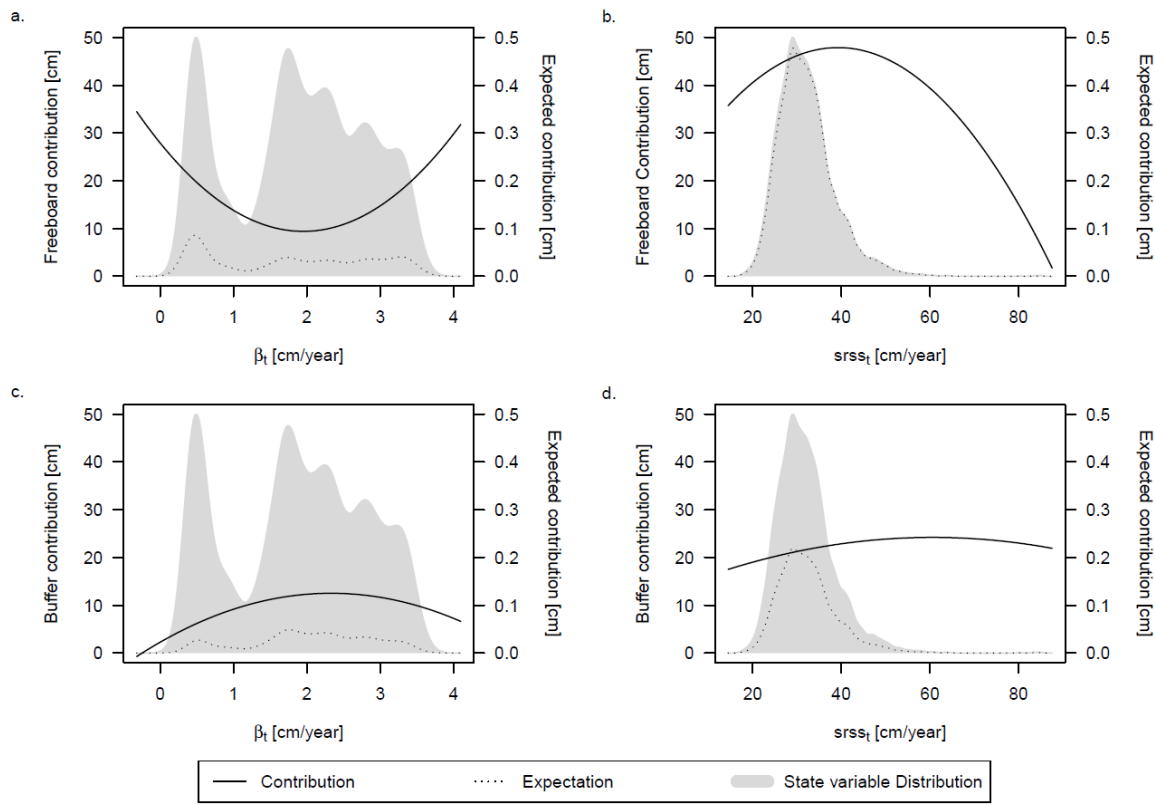

Figure 5. Mapping of state variables ($\beta_t$ - a., c.; $srss_t$ – b., d.) to the freeboard (a., b.) and buffer (c., d.) heights for the minimum total cost solution from the DPS formulation. The black line maps the contribution of the state variable to the appropriate height (left y-axis). The gray polygons are the empirical distributions of the state variables over all SOWs. The dotted line represents the expectation of the contribution given the distribution of the state variables (i.e. black line times the gray polygon; right y-axis).

There is a complex relationship among the parameters that define the SOWs that lead to the intertemporal formulation outperforming the DPS formulation for the minimum total cost solution. We were unable to identify a generalized rule for when this occurs; however, the marginal distributions of the SOW parameters indicate a shift in the shape parameter ($\xi$) and, to a lesser extent, the location parameter ($\mu$) of the surge model as well as the c* parameter associated with the SOWs that favor the intertemporal formulation over the DPS formulation (Fig. 6). These shifts are consistent with SOWs that have a high probability of large and abrupt changes in water level, either due to a threshold event or the storm surge characteristics, from one simulation year to the next.

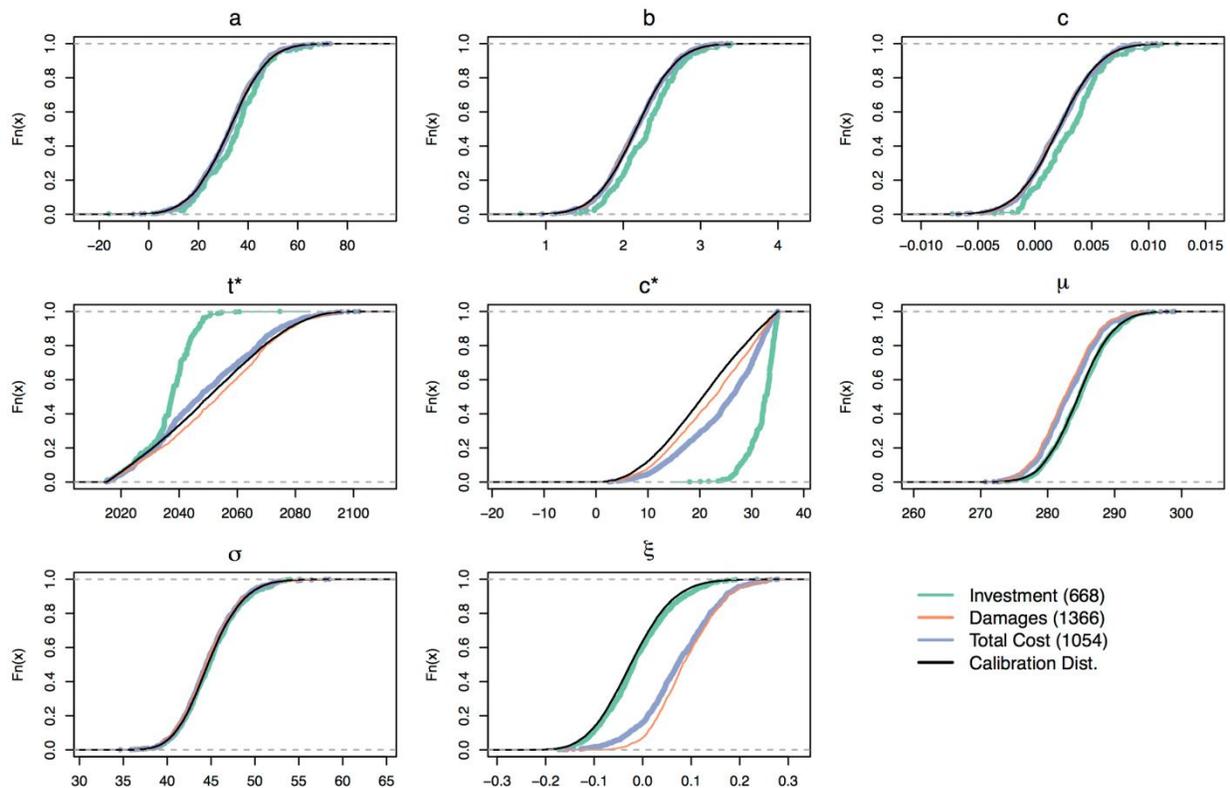

Figure 6. Marginal empirical cumulative distribution functions of SOW parameters from Eqs. 5 and 6 as calibrated in Oddo et al. (2017) (black). Each of the color lines are the distributions of the SOW parameters that lead to the intertemporal formulation producing lower investment cost (green), lower expected damages (orange), and lower total costs (blue) than the DPS formulation for the respective minimum total cost solutions. Numbers to the right of the categories in the legend are the number of SOWs that define the distribution.

## 4. DISCUSSION AND CONCLUSION

We demonstrate that a revised problem formulation that leverages environmental observations can drastically reduce the total social costs of a coastal adaptation problem while increasing the robustness of the fortification to uncertain sea-level rise and storm surge. The DPS formulation, which adapts actions to an observed state of the system, can produce solutions with simultaneously lower expected investment costs and expected damages relative to a more traditional intertemporal approach in far fewer function evaluations. Finding high-quality solutions quickly can open the door to more complex hypothesis testing, such as interactions with other mitigation/adaptation strategies and broader hydrologic system management through reallocation of development and computation time.

The selection of state variables can be vital to the DPS approach. We derive the state variables used in our analysis from a relatively easy-to-observe quantity (maximum annual water level); nevertheless, these state variables provide enough of a signal about the current state of the world for the decision rule to adapt accordingly. More complex derivations of state variables could shift the Pareto-optimal solutions even closer to the ideal solution and better inform SLR adaptation decisions.

While a DPS formulation yields superior solutions relative to the intertemporal formulation, SOWs that have abrupt changes in water level from year to year (i.e. a large surge event or a threshold event that greatly accelerates SLR) can cause the DPS formulation to fail. Our analysis suggests that a great economic value of information may be associated with learning more about the distribution of extreme surge events and the detection of potentially irreversible threshold events that can lead to abrupt SLR. Gaining a better understanding of these deeply uncertain consequences can directly improve the way decisions are made with respect to SLR adaptation.

The results and discussion in this analysis are dependent upon a number of key assumptions and simplifications. First, the base model used in this analysis assumes a single-standing homogenous dike ring. In reality, a single dike is part of a larger hydrologic management system that interacts with other system components. Second, the build policies from both formulations assume that once a decision is made, the heightening is done within the same year. It often takes years to make a decision once the evidence is presented and years more to put the decision to action. Furthermore, with the exception of a dike failure, decision makers do not expect to revisit the dike-heightening situation for several years once the dike has been heightened. While the structure of the model used in this analysis discourages heightening in consecutive years, we do not explicitly account for this constraint. Third, this analysis assumes the economic parameters, such as those related to the investment costs and damages, are known and certain. In many cases, these parameters are deeply uncertain. As a last example, this analysis assumes that the state variables are perfectly observable and observed with enough time to analyze and act accordingly. We used state variables derived from the height of the water against the dike, which is fairly straight-forward to measure. In other climate adaptation decisions, there may be situations where the appropriate state variables needed to inform the decisions are not easy to observe, cannot be observed in time for action, or cannot be observed at all.

In our analysis, we use specific structural models for uncertain SLR and storm surge that were calibrated to local SLR projections and tide gauge data. Additionally, the SLR model allowed for abrupt changes in SLR rates that were uncertain in both onset and magnitude. This treatment of uncertainty allows us to extensively sample a single SLR distribution and a single storm surge distribution; however, both SLR and storm surge are deeply uncertain and this treatment can limit our insights to the real risk surrounding coastal defense infrastructure. As such, the formulation used in this analysis could be complemented by the robust decision making framework to assess the performance of the adaptation solutions to higher-order uncertainties.

We use a discount rate of 4% in this analysis, which is broadly consistent with discount rates used in integrated assessment modelling and cost-benefit analyses of climate change (Adler et al., 2017; Anthoff et al., 2009; Nordhaus, 2017). The use of discount rates in climate-change related decision problems, where action now offsets future damages, is highly-debated (Adler et al., 2017; Heal and Millner, 2014; Nordhaus, 2007; Stern, 2008). Furthermore, we assign the discount rates for both investment costs and damages to the same value. Applying a lower discount rate overall or applying a lower discount rate for future damages relative to the investment cost (i.e. risk-premium) would tend to lead to quicker and/or more aggressive adaptation solutions. Investigating the sensitivity of these results to the choice of discount rates and risk-premiums would be an informative expansion on this work.

Coastal defense infrastructure is required to meet or exceed strict reliability standards. In our analysis, we use a relatively low reliability constraint of 80% for reasons described in section 2.5. Applying a more realistic reliability constraint (e.g. 99% - 99.99%) would make finding feasible solutions more difficult for both formulations (more so for the intertemporal formulation than the DPS formulation). For the sake of our experiment, finding feasible solutions was vital to comparing the two formulations; however, additional analysis with a more strict reliability constraint would provide valuable insight into the viability and practicality of solutions provided by each formulation.

Many of the caveats described above provide avenues to expand this work. This analysis, however, illustrates the need for particular attention to the formulation of the overall problem. SLR and storm surge are deeply uncertain consequences of a changing climate. Using the intertemporal approach and taking expectations over these uncertain SOWs could be leaving valuable information unused. The DPS approach uses this information to learn about the state of the system and determine the appropriate action. The result is a solution set that Pareto-dominates the traditional approach and is more robust to uncertainty. The speed in which the DPS approach converges to Pareto-optimal solutions affords time and resources that can be spent analyzing more complex systems and testing alternative hypotheses. Our analysis demonstrates that multi-objective adaptive formulations can provide important insights for the design of coastal adaptation problems and can identify improved strategies.

**ACKNOWLEDGEMENTS**

The authors would like to thank Perry Oddo, Alex Bakker, Julie Quinn, Pat Reed, Michael Oppenheimer, Vivek Srikrishnan, Tony Wong, and Ben Lee for valuable input.

Funding: This work was partially supported by the National Science Foundation through the Network for Sustainable Climate Risk Management (SCRiM) under NSF cooperative agreement GEO-1240507 as well as the Penn State Center for Climate Risk Management. Any opinions, findings, and conclusions or recommendations expressed in this material are those of the authors and do not necessarily reflect the views of the National Science Foundation.

Author contributions: G.G. and K.K. designed the research. G.G. performed the research, analyzed the results, and wrote the initial draft of the manuscript. G.G. and K.K. proofread and finalized the manuscript.